\documentclass[12pt,a4paper]{conference}

\usepackage{fancyhdr}
\usepackage{graphicx,amsmath,amssymb,cite}
\usepackage{multind}
\makeindex{author} \makeindex{subject}

\newcommand{\beq}{\begin{equation}}
\newcommand{\eeq}[1]{\label{#1}\end{equation}}
\newcommand{\eeqn}{\end{equation}}

\newcommand{\beqa}{\begin{eqnarray}}
\newcommand{\eeqa}[1]{\label{#1}\end{eqnarray}}
\newcommand{\eeqan}{\end{eqnarray}}

\let\bar=\overbar

\newcommand{\Dslash}{\not{\hbox{\kern-4pt $D$}}}
\newcommand{\dslash}{\not{\hbox{\kern-2pt $\del$}}}

\newcommand{\msb}{{\bar{\ssstyle M \kern -1pt S}}}

\begin{document}

\Chapter{On a possible origin of a resonance-like structure in the
two-photon invariant mass spectrum of the reaction $pp \to pp
\gamma \gamma$.}
           {On a possible origin}{A.S.Khrykin \it{et al.}}
\vspace{-6 cm}\includegraphics[width=6 cm]{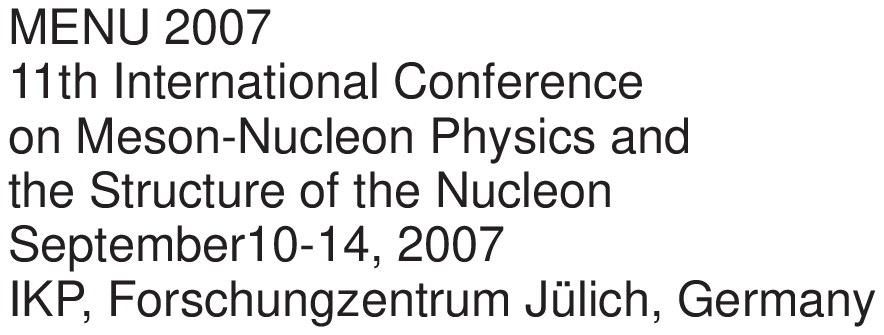}

\vspace{4 cm} \addcontentsline{toc}{chapter}{{\it N. Author}}
\label{authorStart}
\begin{raggedright}
{\it A.S.Khrykin}$^{\star}$$^,$\footnote{khrykin@nusun.jinr.ru }~,
{\it S.B.Gerasimov}$^{\#}$\\
\bigskip
$^{\star}$Dzelepov Laboratory of Nuclear problems, Joint Institute for Nuclear Problems, 141980 Dubna, Moscow Region, Russia\\
$^{\#}$Bogolubov Laboratory of Theoretical Physics, Joint Institute for Nuclear Problems, 141980 Dubna, Moscow Region, Russia\\
\end{raggedright}
\begin{center}
\textbf{Abstract}
\end{center}
We show that the resonance-like structure found by the
\emph{CELSIUS-WASA Collaboration} in the two-photon invariant mass
spectrum of the reaction $pp \to pp \gamma \gamma$ is rather a
signature of the $NN$-decoupled dibaryon resonance
$d^\star_1$(1956) that is produced in the radiative process $pp
\to \gamma d^\star_1$ and then undergoes radiative decay into two
protons $d^\star_1 \to pp \gamma$. It is found that a contribution
of the dibaryon mechanism $pp \to \gamma d^\star_1 \to pp \gamma
\gamma$ of the reaction $pp \to pp \gamma \gamma$ to the invariant
mass spectrum of its photon pairs can reasonably well reproduce
the experimentally observed spectrum in the vicinity of the
resonance-like structure.
\section{Introduction}
 The resonance-like structure found by the \emph{CELSIUS-WASA
Collaboration} in the two-photon invariant mass spectrum of the
exclusive reaction $pp \to pp \gamma \gamma$ at 1.2 and 1.36 GeV
has been taken by the authors of Ref. \cite{CELWAS} as evidence
for dynamical formation of the S-wave dipion resonance $\sigma$
\cite{BrownSinger} in the $pp$ collision process. It was assumed
\cite{CELWAS}  that this structure might result from interference
between the process $pp \to pp \sigma \to pp \gamma \gamma$ and
one of the double $pp$-bremsstrahlung. However, such an
interpretation is at least questionable merely because the
amplitude of the double $pp$-bremsstrahlung  is unknown in the
energy range considered in Ref.\cite{CELWAS} where this process,
to our knowledge, has not been investigated yet either
experimentally or theoretically. The aim of this paper is to
propose an alternative interpretation of a possible origin of this
structure which is based on the dibaryon mechanism of the
two-photon emission in $NN$ collisions \cite{PRC64, NPA721}.
\section{The dibaryon mechanism of two-photon emission in $NN$ collisions}
The dibaryon mechanism of two-photon emission in $NN$ collisions
$NN \to \gamma d^\star_1 \to NN \gamma \gamma$ governs the
electromagnetic transition between the initial and final $NN$
states by a sequential emission of two photons, one of which is
caused by production of the $NN$ decoupled dibaryon resonance
$d^\star_1$ and other by its subsequent decay. In the overall
center-of-mass system the energy of the photons $E^F_\gamma$
associated with the $d_{1}^{*}$ production is determined by the
dibaryon mass $M_R$ and the energy of colliding nucleons
$W=\sqrt{s}$ as $E^F_\gamma=(W^2-M_R^2)/{2W}$. The energy of
photons $E^D_\gamma$ emerging as a result of the $d_{1}^{*}$ decay
in the resonance rest frame is given by
$E^D_\gamma=(M_R^2-M_{NN}^2)/2M_R$, where $M_{NN}$ is the
invariant mass of the final $NN$ state. The matrix element for the
$NN \to NN \gamma \gamma$ transition will in general be a function
of the four-momenta of the incoming and outgoing particles
together with the mass and quantum numbers of the resonance
$d^\star_1$ which are still not established. It can be written in
the form $\mathcal{M}=\mathcal{M}_F \cdot \mathcal{D}(p_R) \cdot
\mathcal{M}_D,$ where $\mathcal{M}_F$ and $\mathcal{M}_D$ are the
matrix elements for the dibaryon formation and decay,
$\mathcal{D}(p_R)=1/({p_R^2-M_R^2+iM_R{\it{\Gamma_R}}})$ is the
propagator of the dibaryon with the four-momentum $p_R$ and
${\it\Gamma_R}$ is its decay width. In this work the $pp \to pp
\gamma \gamma$ transition has been treated within the assumption
that at large distances the $NN$-decoupled six-quark $d_{1}^{*}$
state is a bound $p\Delta(1232)$ state with the spin-parity
$J^P=0^-$ and the isospin $I=2$ \cite{Mats'97}. Owing to a
relatively small energy $\sim$ 80 MeV released in the $d_{1}^{*}$
decay, the matrix element $\mathcal{M}_D$ was derived in terms of
a simple picture in which the decay $(p \Delta)_{bound}\to
pp\gamma$ proceeds via the virtual $\Delta^{+}\to p\gamma$
$M1$-transition. As a radial wave function for the bound
$p\Delta$-state we have considered two functional forms: the
Gaussian $R_{G}(r)=N_{G}\cdot r\exp(-b^{2}_{G}r^2)$ and the
Fermi-type (or Woods-Saxon) distribution
$R_{F}(r)=N_{F}\cdot(1-j_{0}(\kappa_{F} r))$ for $r < R_{0}$ and
$R_{F}(r)=N_{F} \cdot C_{F}\cdot [1+\exp[a\cdot(r-R_{0})]]^{-1}$
for $r\geq R_{0}$, where $r=|\vec{r}_{\Delta}-\vec{r}_{p}|$,
$a=\sqrt{2m_{red}E_{b}(d_{1}^{*})}$, $m_{red}$ is the reduced mass
of the $(p\Delta)$-system, $E_{b}=M_{\Delta}+m_{N}-M_R$ is the
binding energy of this system, $N_{F(G)}$ are the normalization
constants, the parameters $C_{F}$, $R_0$ and $\kappa_{F}$ are
defined from the continuity requirements for $R_{F}(r)$ and its
first and second derivatives at $r=R_{0}$. The specific form of
the correlation function $f(r)=1-j_{0}(\kappa_{F} r)$ ($j_0(z)$ is
the spherical Bessel function of $0^{th}$ order), describing
effects of the soft core in the $N\Delta$-interaction potential is
taken in accordance with the Ref.\cite{Weise'77}. The parameter
$b^{2}_{G}$ is chosen such that the
rms radius of the $(p \Delta)$-state is the same for both versions of the wave function. \\
Unlike the decay process, the $d_{1}^{*}$ formation one takes
place at relatively high energies of colliding protons. Therefore,
its mechanism may be more involved. The lack of an explicit theory
of such a process forced us to resort to the phenomenology.
Namely, we adopt $|\mathcal{M}_F|^2 c.m. \simeq A\cdot
exp(-k_\perp/b),$ where $A$ is the normalized constant, $k_\perp$
is the transverse momentum of a photon and $b$ is a parameter.
This formula was shown \cite{And'71} to give a good fit for the
reaction $pp \to \pi^{+} d$ for incident proton momenta in the
range $3.4 - 12.3$ GeV/c with $k_\perp \rightarrow
p_\perp(\pi^{+})$. According to \cite{ALLABY}, it is equal to 0.26
GeV/c at $E_{c.m.}=3.0$ GeV. In our calculation we used $b=0.6$
CeV/c. This value follows from the assumption that $b$($pp\to
\gamma d^\star_1$)/$b$($pp \to \pi^{+} d$)=rms radius($d$)/rms
radius($d^\star_1$).\\
 The effects of the final state interactions
between decay protons in the $^{3}P_{1}$-state were included with
the help of the phenomenological correlation function
$f_{phen}(r)=1-j_{0}(\kappa r), \kappa=3.93~fm^{-1}$ by
multiplying the $L=1$ radial wave function of free motion by this
function. The approximate relevance of this procedure is
demonstrated numerically in \cite{Par'02}.
\section{The method of calculations and results}
The calculations of the invariant mass spectra of photon pairs
from the dibaryon mechanism of the reaction $pp\to pp
\gamma\gamma$ at 1.36 GeV for the geometry and kinematics of the
experiment \cite{CELWAS} were done using the Monte Carlo method. A
computer program for the MC calculations was made on the basis of
the GENBOD event generator \cite{GENBOD} which was used to
randomly generate four-momenta of particles for the process $pp
\rightarrow \gamma + d^\star_1 \rightarrow \gamma+ \gamma+pp$. A
probability of any event was given by its weight
$WT=`\Sigma_{\sigma_{1,2,3,4}}\Sigma_{\lambda_{1,2}}|\mathcal{M}|^{2}$,
where $\sigma_{i}$~are the spin projections of the protons and
$\lambda_{i}$ stand for the polarizations of the photons. The
calculated spectra for both forms of the radial wave function for
the bound $p\Delta$-state are presented in Fig.1. Both spectra are
seen to reproduce reasonably well the experimentally observed
spectrum \cite{CELWAS} in the vicinity of the resonance-like
structure. The spectrum calculated with the matrix element from
Ref.\cite{Scholten}, which was obtained for the case of point-like
dibaryon, is also shown in the same figure for comparison. In this
calculation the effects of the final state interactions were ignored.\\
Our results show that the structure observed in the two-photon
invariant mass spectrum of the reaction $pp \to pp\gamma\gamma$ by
the CELSIUS-WASA collaboration is very likely to be due to the
dibaryon mechanism of the reaction $pp\to pp
\gamma\gamma$\cite{PRC64, NPA721}. They can thus be considered as
one more confirmation of the existence of this two-photon
production mechanism in $NN$ collisions and, hence, the existence
of the dibaryon $d^\star_1$ itself. In this connection we note
that more experimental and theoretical studies are needed to
completely clarify the situation with the existence of the
dibaryon resonance $d^\star_1$.
\begin{figure}[htb]
\begin{minipage}[c]{75mm}
\includegraphics[width=75 mm]{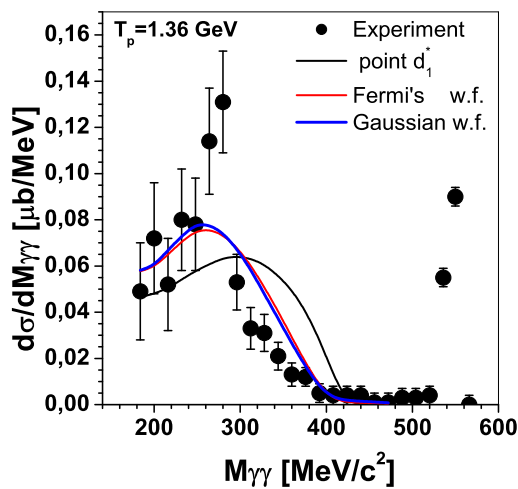}
\end{minipage}
\hspace{\fill}
\begin{minipage}[c]{75mm}
 \caption{\small{Experimentally observed two-photon
invariant mass spectrum of the reaction $pp\gamma\gamma$ and those
for the process $pp \to \gamma d^\star_1 \to \gamma\gamma pp$
calculated with two different types of radial wave function of the
bound $p\Delta$-state, Fermi's and Gaussian. The spectrum
calculated with the matrix element from \cite{Scholten} is given
by the black solid line.}}
\end{minipage}
\end{figure}


\begin{thebibliography}{0}    
\bibitem{CELWAS} M. Bashkanov, H. Clement, E. Doroshkevich, M.
Kaskulov, R. Meier, T. Skorodko and G.J. Wagner, Int. Jour of Mod.
Phys. {\bf A20}, 554(2005);hep-ex/0406081.
\bibitem{BrownSinger} L. M. Brown and P. Singer, Phys. Rev.
Letters {\bf 8}, 460 (1962).
\bibitem{PRC64} A. S. Khrykin et al., Phys. Rev. C {\bf 64}, 034002(2001).
\bibitem{NPA721} A. S. Khrykin, Nucl.Phys. {\bf A721}, 625c(2003).
\bibitem{Mats'97} A. Matsuyama, Phys. Lett. {\bf B 408}, 25 (1997).
\bibitem{Weise'77} W. Weise, Nucl.Phys. {\bf A278}, 402 (1977).
\bibitem{And'71} H. L. Anderson, et al., Phys. Rev. D {\bf 3}, 1536 (1971).
\bibitem{ALLABY} J.V. Allaby et al., Phys. Lett. {\bf B 29}, 198 (1969).
\bibitem{Par'02} A. Parreno and A. Ramos, Phys. Rev. {\bf C 65}, 015204(2002), arXiv:nucl-th/0104080
\bibitem{Scholten} O. Scholten, J. G. O. Ojwang, and S. Tamenaga, Phys. Rev. C {\bf 71}, 034005(2005).
\bibitem{GENBOD} F. James, \emph{Monte Carlo Phase Space}, CERN
68-15. 1968.
\end{thebibliography}
\end{document}